\documentstyle[prl,aps,epsfig]{revtex}

\begin{document}
\twocolumn[
\hsize\textwidth\columnwidth\hsize\csname@twocolumnfalse\endcsname

\title{Critical exponents in Ising spin glasses} \author{L.W. Bernardi and
I.A. Campbell}

\address{Laboratoire de Physique des Solides,\\ Universit\'e Paris Sud,
91405 Orsay,
France}

\maketitle

\begin{abstract}
We determine accurate values of ordering temperatures and critical
exponents for Ising Spin Glass transitions in dimension 4, using a
combination of finite size scaling and non-equilibrium scaling techniques.
We find that the exponents $\eta$ and $z$ vary with the form of the
interaction distribution, indicating non-universality at Ising spin glass
transitions. These results confirm conclusions drawn from numerical data
for dimension 3.
\end{abstract}
\vfill
 \twocolumn
\vskip.5pc ]
\narrowtext

\section{Introduction}
Precise values of critical exponents at Ising Spin Glass (ISG) transitions
have been hard to obtain from numerical simulations principally because of
the agonizingly slow relaxation near the transition. Finite size 
scaling~\cite{1} can help bypass this problem but can introduce other 
difficulties, associated with the possible need to allow for corrections to 
the finite size scaling rules. We have shown~\cite{2} that in favourable cases 
a combination of finite size scaling and non-equilibrium scaling techniques
can  allow one to estimate to high accuracy both the ordering temperature
and the exponents $\eta$ and $z$. We exploited this method for ISGs in
dimension 3 where we found that the exponents $\eta$ and $z$ changed
considerably as a function of the distribution of near neighbour
interactions, in complete contradiction to what would be expected from
standard second order universality rules. There is no formal theorem which
states that ISG transitions should obey universality rules, but up to now
general sentiment tended to be in favour of universality. The empirical
observation of non-universality~\cite{2} implies that the standard
application of the renormalization group method to ISGs must be
reconsidered, and shows that there are important qualitative differences
between ISG transitions and conventional second order transitions.

We have now applied the same approach to ISGs in dimension 4 and for one
case in dimension 5. We find that in dimension 4 as in dimension 3 the
exponents $\eta$ and $z$ vary systematically from system to system. The
numerical results for the exponents as functions of dimension can be
compared with the predictions of the $\epsilon$ expansion; agreement is
very poor if the expansion is extended beyond the leading term.

\section{Method}

Technically the problem of measuring exponents is less arduous in higher
dimensions than in dimension 3. Because dimension 4 is well above the lower
critical dimension the existence of a finite $T_g$ in dimension 4 has never
been in doubt, in contrast to the situation in dimension 3. The Binder
cumulant curves for each system show clear intersections and there already
exist quite precise estimates of the ordering temperatures $T_g$, at least
for the $\pm J$ and Gaussian interactions~\cite{3,4,5,6,7}.

We will first outline the simulation techniques that we have used. First we
measure, for a number of test temperatures $T$ near the probable value of
$T_g$, the time dependent non-equilibrium spin glass susceptibility
$\chi^*(t)$. Two completely random (infinite temperature) replicas $A$ and
$B$ of the same system are quenched to $T$ with independent updating. The
spin glass susceptibility
\begin{equation}
\chi^*(t)= [<S_i^A(t)S_i^B(t)>^2]
\end{equation}
is recorded as a function of time $t$ after the start of the quench.
Precisely at $T_g$, $\chi^*$ increases with $t$ as $t^h$, with
$h=(2-\eta)/z$~\cite{8}. If we do not know $T_g$ a priori, we obtain an
apparent $h(T)$ at each test temperature $T$. This non-equilibrium
parameter presents the considerable advantage of requiring no preliminary
anneal, and represents the growth of correlations as the internal
temperature drops in the ISG samples. Because we concentrate on short times
after having chosen an appropriately sized sample the correlation lengths
are always much smaller than the sample size so there are no finite size
corrections to a very good approximation. This type of non-equilibrium
scaling has been tested very carefully on regular systems where the
ordering temperature and the exponents were already known~\cite{9}, and the
scaling has been rigorously verified.

Secondly, at the same test temperatures $T$ we anneal samples for a waiting
time $t_w$ before measuring the initial decay of the autocorrelation
function over a further time~$t$
\begin{equation}
q(t) = <S_i(t_w)S_i(t+t_w)>
\end{equation}
If $t<<t_w $ and $T = T_g$, $q(t) =t^{-x}$ with $x=(d-2+\eta)/2z$
\cite{10}. Under these conditions the decay of $q(t)$ is identical to the
equilibrium form of the decay (which would be measured after an infinitely
long anneal time), and again if the sample size is chosen appropriately
there are no finite size corrections~\cite{11}. It turns out that that at
temperatures $T<T_g$, $q(t)$ follows an algebraic decay law with effective
values of $x$ which are smaller than the value corresponding to $T=T_g$.
For higher temperatures there is a further multiplicative factor
$f(t/\tau)$.

It can be seen that if reasonable precautions are taken the effective
exponent combinations $h(T)$ and $x(T)$ can be measured with negligable
systematic error. One can thus obtain a first set of effective exponents
$\eta_1(T)$ and $z(T)$ as functions of the test temperature $T$ from the
equations
\begin{equation}
z(T) = \frac{d}{2x+h}
\label{eq:z}
\end{equation}
\begin{equation}
\eta_1(T) = \frac{4x-h(d-2)}{2x+h}
\end{equation}

Finally we use finite size scaling of the equilibrium spin glass
susceptibility as a function of sample size $L$. Again at $T_g$ the
normalized spin glass susceptibility $\chi_{SG}/L^2$ is proportional to
$L^{-\eta}$~\cite{1} giving an independent measure of $\eta$. Below $T_g$
the power law form continues to hold, providing a second measure of $\eta$,
$\eta_2(T)$. Now if we plot $\eta_1(T)$ and $\eta_2 (T)$ against $T$,
consistency dictates that the true $T_g $ and $\eta$ must correspond to the
intersection point of $\eta_1 (T)$ and $\eta_2 (T)$.

\section{Results in dimension 4}

We have made measurements for ISGs on the 4 dimension simple cubic lattice,
with near neighbour interactions of different types. In practice the finite
size scaling is the most demanding part of the simulation as far as CPU
hours are concerned, as it is essential to make stringent checks that
equilibrium has been achieved, and averages must be made over a large
number of samples. We have been fortunate in being able to use equilibrium
$\chi_{SG}$ data from A.P. Young~\cite{3} for the 4d $\pm J $ interaction
case and from Parisi et al~\cite{5} for the 4d Gaussian interaction case.
For the other two cases we carried out simulations in the usual manner for
$L$ up to 8 taking standard precautions~\cite{1} to be certain of complete
thermal equilibrium at each size.

We thus dispose of data for the 4 dimension $\pm J$ (J), Uniform (U),
Gaussian (G), and decreasing exponential (Ed) interactions. Earlier work
\cite{3,4,5,6,7} had shown that $T_g$ is close to 2.0 for the $\pm J$ case
and close to 1.75 for the Gaussian case. To get a trial estimates of $T_g$
for the Ed and U cases we have relied on a Migdal-Kadanoff approach~\cite{12} 
where it is assumed that the normalization scaling parameter $b$
is the same for all members of a family of ISGs in a given dimension. This
gives quite reliable values of $T_g$ once the value for one member of the
family is known. Simulations for $h(T)$ were made using up to 2000 samples
of size $L= 10$. $x(T)$ measurements were made on samples of size $L = 20$
with anneals of $10^6$ MCS and runs of $10^4$ MCS. Runs were taken with one
sample.

\begin{figure}
\begin{center}
\epsfig{file=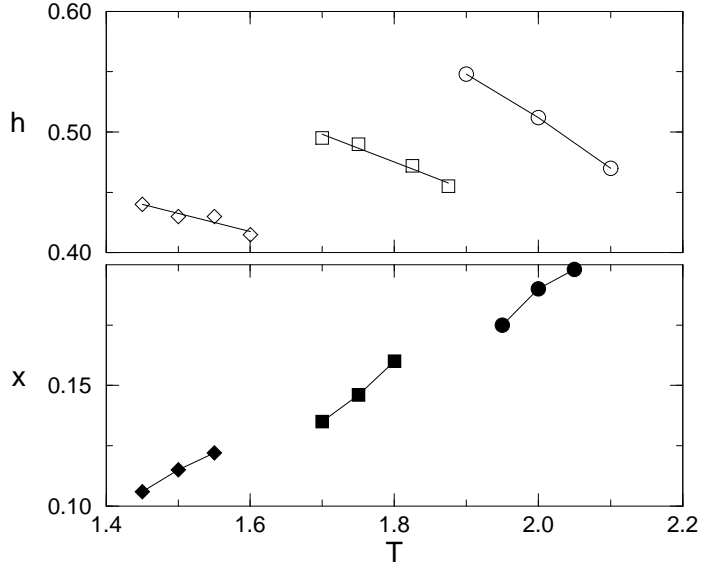,width=8.9cm}
\end{center}
\caption{The effective $h(T)$ and $x(T)$ (see text) as functions of
temperature $T$ for the $\pm J$ interaction ($\circ$), the Gaussian
interaction ($\Box$) and the decreasing exponential interactions
($\Diamond$). Full symbols: $x$. Empty symbols: $h$.}
\label{fig:1}
\end{figure}

Data points for $h(T)$ and $x(T)$ are shown in figure~\ref{fig:1}. 
It can be seen that for each system the effective value $x(T)$ increases with 
$T$ while the effective value $h(T)$ decreases with $T$. In dimension 3, $h(T)$
increases with $T$~\cite{2}; we have no explanation for this difference.

\begin{figure}
\begin{center}
\epsfig{file=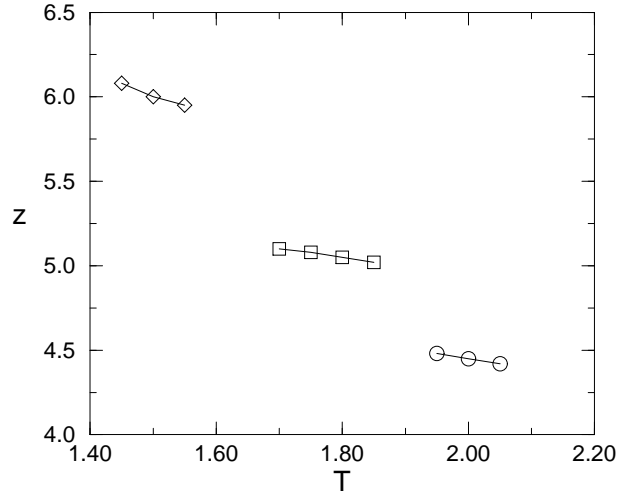,width=8.9cm}
\end{center}
\caption{The effective dynamic exponent $z$ obtained using
equation~\ref{eq:z} for $\pm J$ interaction, gaussian interaction and
decreasing exponential interactions. (same symbols as figure~\ref{fig:1}).}
\label{fig:2}
\end{figure}
We first extract $z(T)$ from the $h(T)$ and $x(T)$ data using 
equation~\ref{eq:z}, figure~\ref{fig:2}.  Clearly $z$ is not
universal; $z$ tends to increase steadily from one type of system to the
next as the kurtosis of the interaction distribution increases and $T_g$
drops.

\begin{figure}
\begin{center}
\epsfig{file=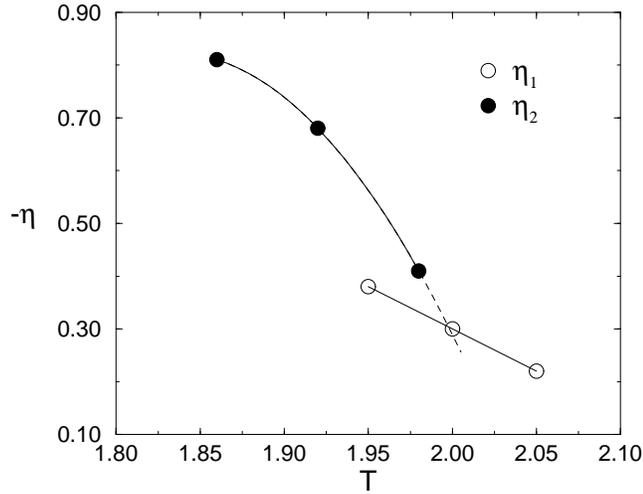,width=8.9cm}
\end{center}
\caption{Effective exponents $\eta_1(T)$ ($\circ$) and $\eta_2(T)$ ($\bullet$)
(see text) as function of temperature for the $\pm J$ interaction case. The 
intersection gives $T_g$ and the exponent $\eta$.}
\label{fig:3}
\end{figure}

In figure~\ref{fig:3} we have plotted $\eta_1(T)$ and $\eta_2 (T)$ for the
$\pm J$ case where we have the most complete $\eta_2$ data.
There is a clear intersection at the point $T_g = 2.00 \pm 0.01$ and 
$\eta =-0.30 \pm 0.02$. High quality Binder cumulant data for $L=4$ to 12 
due to Young~\cite{3}
show a clearly defined crossing of all the $g_L(T)$ curves at $T =2.00 \pm
0.01$~\cite{14}. Two series expansion calculations gave $T_g$ estimates of
$2.02 \pm 0.06$~\cite{6} and $2.04 \pm 0.05$~\cite{7}. The agreement
between different determinations of $T_g$ in this case is thus very
satisfactory, giving further confidence in the new method outlined above.
We can note that for estimating $\eta$, this method is fairly insensitive
to the exact value of $T_g$ because the line for $\eta_1(T)$ is much less
steep in the region of $T_g$ than is the line for $\eta_2(T)$. Also the
present method estimates $\eta$ at the ordering temperature only, and so
does not imply any assumption of a scaling relation over a range of
temperatures around $T_g$.

\begin{figure}
\begin{center}
\epsfig{file=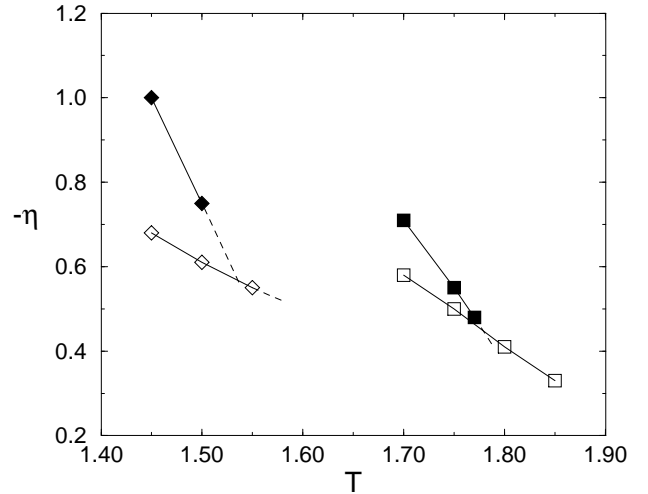,width=8.9cm}
\end{center}
\caption{As in figure~\ref{fig:3}, for the Gaussian interaction case
($\Box$) and the decreasing exponential case ($\Diamond$). Empty symbols:
$\eta_1$. Full symbol: $\eta_2$} 
\label{fig:4}
\end{figure}

If we now turn to the other sets of interaction distributions, 
Figure~\ref{fig:4}, we find similar intersections leading to the $T_g$ and 
$\eta$ values given in Table~\ref{tab:1}. 

The $T_g$ value in the Gaussian interaction case lies between
the two published estimates obtained from Binder cumulant work, and is
close to estimates we can obtain starting from the $\pm J$ value for $T_g$
and using the Migdal-Kadanoff technique~\cite{12} or by using the series
expansion formula due to Singh and Fisher~\cite{15} (both methods give
essentially $T_g =1.75$ for the Gaussian case if $T_g$ is taken to be 2.00
for the $\pm J$ case).

 For the U and Ed cases there are no Binder cumulant
data to compare with but the Migdal-Kadanoff and Singh-Fisher techniques
lead to $T_g$ values of 1.90 and 1.53 respectively for the two interaction
distributions. Overall agreement between the simulation estimates and these
values can again be seen to be excellent. The $\eta$ values, 
Table~\ref{tab:1}, like the $z$ values, are strikingly non-universal, with 
$\eta$ becoming more negative as the kurtosis of the distribution increases 
and the $T_g$ drops.

\section{Dimension 5}
For the $\pm J$ interaction ISG in dimension~5, series expansion~\cite{7}
gives a precise value for the ordering temperature, $T_g = 2.57 \pm 0.01$.
We have seen above that in dimension 4 the simulations and the series work
led to very similar $T_g$ estimates, so the dimension 5 series estimate
should be very reliable. We have asumed this $T_g$ value is correct and
have measured $h$ and $x$ at this $T_g$. The values, Table~\ref{tab:1}, lead 
to  $z=4.50 \pm 0.1$ and $\eta=-0.39 \pm 0.02$, in excellent agreement with 
the series estimate $\eta=-0.38\pm 0.07$.

We have also measured $h$ for the Gaussian ISG in dimension~5 ($T_g=2.31$),
and for the $\pm J$ ISG in dimension~6 at $T_g$ which is equal to 
3.03~\cite{7,16}.

\section{Critical exponents - dimensional dependence.} 

We can first concentrate on the behaviour of the critical exponents and their
combinations as a function of dimension for the series of ISGs with $\pm J$
interactions, in order to compare with $\epsilon$ expansion 
expressions~\cite{17,18}. In the standard renormalisation group treatment of 
the ISGs, the upper critical dimension is 6 and calculations have been made 
to third order in $\epsilon$ ($\epsilon = 6-d$).

\begin{figure}
\begin{center}
\epsfig{file=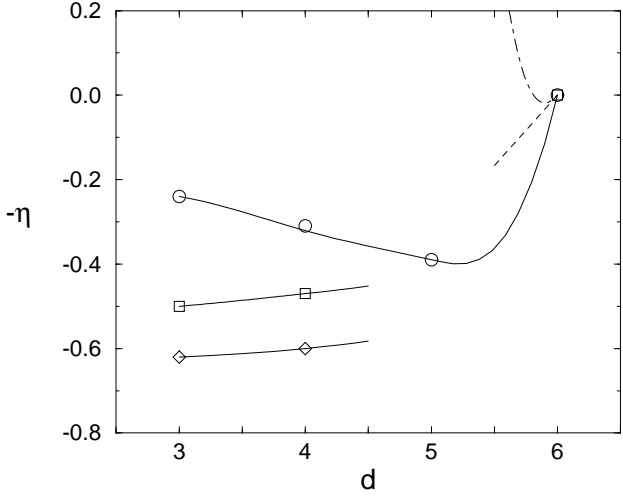,width=8.9cm}
\end{center}
\caption{$\eta$ as a function of dimension for the $\pm J$ interaction
($\circ$), the Gaussian interaction ($\Box$) and the decreasing exponential
interactions ($\Diamond$). The full lines are to guide the eye. The
straight dashed line indicates the $\epsilon$ expansion to first order; the
dot dashed curve is the $\epsilon$ expansion to order 3.}
\label{fig:5}
\end{figure}

1/ $\eta (d)$. The values of $\eta$ from \cite{2} and the present work are
shown in Figure~\ref{fig:5}, with an extrapolation to the upper critical
dimension value $\eta = 0$ for dimension 6.
The renormalisation group $\epsilon$ expansion estimate to leading order in 
$\epsilon$ and to order 3 is also indicated. It can be seen that the 
simulation values, which from the discussion given above we consider very 
reliable, behave in a regular fashion. The initial trend of $\eta(d)$ 
extrapolated towards $d=6$ is broadly consistent with the $\epsilon$ expansion 
to lowest order. However the $\epsilon$ expansion curve to order 3 lies a 
long way from the numerical points. This is in striking contrast to the 
standard second order transition case where the $\epsilon$ expansion to the 
same order gives excellent agreement with numerical or analytic values right 
down to $\epsilon = 2$.

\begin{figure}
\begin{center}
\epsfig{file=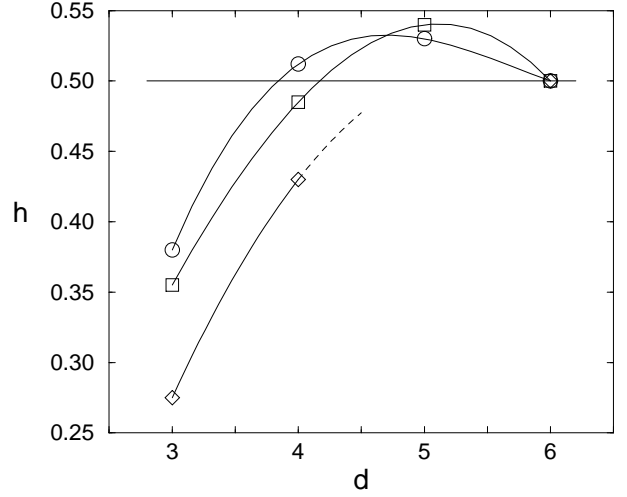,width=8.9cm}
\end{center}
\caption{The exponent $h$ as a function of dimension for the $\pm J$
interaction ($\circ$), the Gaussian interaction ($\Box$) and the decreasing
exponential interactions ($\Diamond$). The point at dimension 6 is
measured.}
\label{fig:6}
\end{figure}

2/ $h(d)$. Using van Hove arguments, Zippelius \cite{18} found that that
below the upper critical dimension $z=2(2-\eta)$, with no correction to
leading order in $\epsilon$. From the definition of $h$, this relation can
be simply rewritten $h=0.5$, so we would expect $h$ to be close to this
value at and near to $d = 6$. Indeed we find that $h$ is equal to 0.5 at $d=6$,
and that $h(d)$ does stay close to 0.5 down to dimension 4, Figure~\ref{fig:6}.

This shows that the Zippelius relation between $z$ and $\eta$ is fairly
accurate.

\begin{figure}
\begin{center}
\epsfig{file=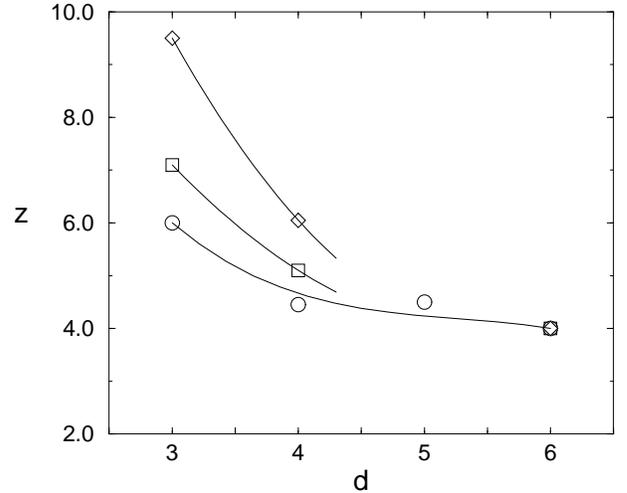,width=8.9cm}
\end{center}
\caption{The dynamic exponent $z$ as a function of dimension for the $\pm
J$ interaction ($\circ$), the Gaussian interaction ($\Box$) and the
decreasing exponential interactions ($\Diamond$). The point at dimension 6
is the theoretical upper critical dimension value.}
\label{fig:7}
\end{figure}

3/ $z(d)$. At dimension 6 we expect $z=4$; the $z$ data as a function of
dimension are approaching 4 as dimension is increased towards 6,
Figure~\ref{fig:7}.

\begin{figure}
\begin{center}
\epsfig{file=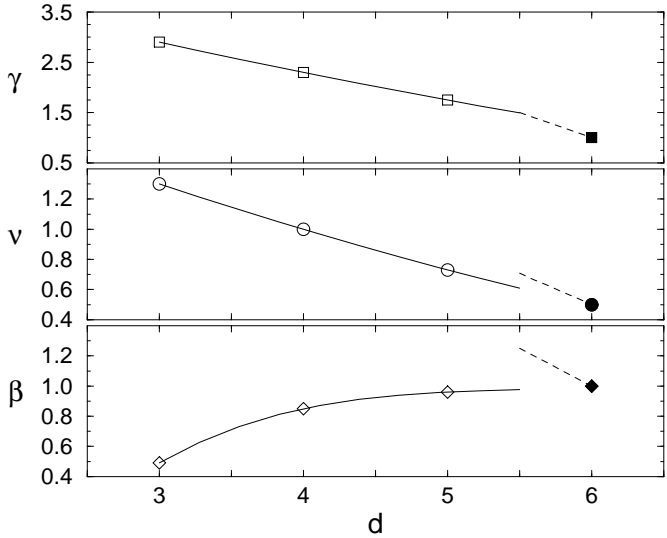,width=8.9cm}
\end{center}
\caption{A collection of critical exponents for the $\pm J$ interaction
case as a function of dimension. The $\nu$ values are taken from references
\protect\cite{3,7,10} and the $\gamma$ and $\beta$ values are
calculated from $\eta$ and $\nu$ using the standard scaling relations. The full
points at $d=6$ correspond to the theoretical upper critical dimension
values and the dashed lines represent the theoretical $\epsilon$ expansions
to first order. }
\label{fig:8}
\end{figure}

4/ $\nu(d)$. We have no new information on $\nu$ from our simulations, but
for completeness we present results compiled from various sources~\cite{3,7,10}
in Figure~\ref{fig:8} together with the $\epsilon$ expansion curve.

Once $\eta$ and $\nu$ are known, the other static exponents can all
be deduced using the scaling rules, Figure~\ref{fig:8}.

\section{Critical exponents - interaction dependence.}

It can be seen in Table~\ref{tab:1} that (as in dimension 3), the exponents in
dimension 4 vary in a systematic manner, with $\eta$ becoming more negative
and $z$ tending to a higher value when the kurtosis of the interaction
distribution increases and the $T_g$ drops. (There are many other possible
parameters which could be used to modify the interaction distribution --
for instance an alternative way to increase the kurtosis would be to dilute
the interactions. We do not know if there is a one to one relationship
between the exponents and the kurtosis in all cases).

\begin{figure}
\begin{center}
\epsfig{file=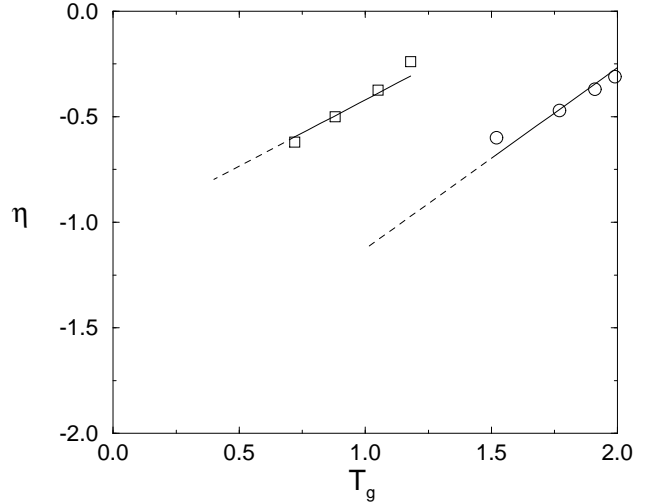,width=8.9cm}
\end{center}
\caption{The trend of $\eta$ as a function of $T_g$ for dimension 3
(circles) and dimension 4 (squares). }
\label{fig:9}
\end{figure}

We can attempt to understand from a na\"{\i}ve standpoint what the
systematic behaviour is indicating~\cite{19}. If we make the heuristic
assumption that the kurtosis is a pertinent parameter as far as the
critical exponents are concerned, then as the kurtosis is increased and the
$T_g$ is driven towards zero, we would expect $\eta$ and $z$ to tend
concomitantly towards their zero $T_g$ values, which are $2-d$ and infinity
respectively. If we plot, Figure~\ref{fig:9}, $\eta$ as a function of $T_g$
for dimensions 3 an 4, we find that the trend of the numerical values is 
consistent with a tendency for $\eta$ towards $2-d$ as $T_g$ tends towards 
zero.

\begin{figure}
\begin{center}
\epsfig{file=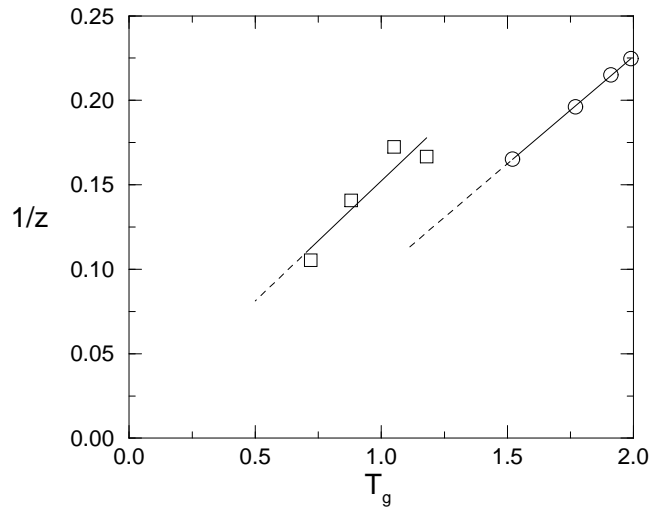,width=8.9cm}
\end{center}
\caption{$\frac{1}{z}$ plotted against $T_g$ for dimension 3 ($\Box$) and
dimension 4 ($\circ$). $\frac{1}{z}$ extrapolates toward zero as $T_g$
decreases, indicating that $z$ diverges when $T_g$ goes to zero.}
\label{fig:10}
\end{figure}

Plotting, Figure~\ref{fig:10}, $1/z$ against $T_g$ we again find a trend 
corresponding to a divergence of $z$ as $T_g$ tends to zero.

We have so far studied the effect of just one parameter, the form of the
interaction distribution, on the critical exponents in finite dimension
ISGs. Other parameters readily suggest themselves; it may well be that
there is a continuous variation of the values of the critical exponents
whenever any parameter is changed, so that the universality class concept
simply does not apply to spin glasses. It is certainly relevant that the
critical exponent $z$ has been shown analytically to change continuously as
a function of applied field along the AT line for the mean field spin 
glass~\cite{20}. It appears that the physics of transitions in complex systems
such as spin glasses is fundamentally different from that of second order
transitions in regular systems.

\section{Conclusion}

We have found accurate values for the the ordering temperatures $T_g$ and
for the critical exponents $\eta$ and $z$ in Ising Spin Glasses with
different sets of near neighbour interactions on the simple cubic lattice
in dimension 4. The results demonstrate that, as in dimension~3~\cite{2},
the exponents are not universal, confirming that the universality class
concept is not relevant to spin glass transitions. This observation shows
that the spin glass transition cannot be treated as a second order
transition in the conventional sense, and that the renormalization group
theory as applied to spin glass transitions (and by extension to
transitions in all complex systems) should be seriously reconsidered.

\section {acknowledgements}
The numerical calculations were carried out thanks to a time allocation
provided by IDRIS ( Institut du D\'eveloppement des Ressources en
Informatique Scientifique). We would like to thank A.P. Young and F. Ritort
for generously providing us with numerical data, which we used to help
construct Figures~\ref{fig:3} and \ref{fig:4}. We profited from
enlightening discussions with D.J.W. Geldart.

\begin{table}[ht]
\begin{center}
 \begin{tabular}{ccccccc} 
d & System  & $T_g$        & $h(T_g)$     &    $\eta$     &   $z$    \\ \hline
4 & $\pm J$ & 1.99$\pm$.01 & 0.51$\pm$.02 & -0.31$\pm$.01 & 4.45$\pm$.1 \\ 
  &  U      & 1.91$\pm$.01 & 0.51$\pm$.02 & -0.37$\pm$.02 & 4.65$\pm$.1 \\   
  &  G      & 1.77$\pm$.01 & 0.48$\pm$.02 & -0.47$\pm$.02 & 5.10$\pm$.1 \\ 
  &  Ed     & 1.52$\pm$.01 & 0.43$\pm$.02 & -0.60$\pm$.03 & 6.05$\pm$.1 \\ \hline
5 & $\pm J$ & 2.57         & 0.530$\pm$.02 & -0.39$\pm$.02 & 4.50$\pm$.1 \\ 
  &  G      & 2.31         & 0.537$\pm$.02 &       ---     &   ---       \\ 
  \end{tabular}
\end{center}
\caption{Values of the temperature of transition and critical exponents for ISG
for various distributions of interactions in dimension 4 and 5. The temperature
of transition in dimension 5 where taken from series expansion}
\label{tab:1}
\end{table} 
\narrowtext


\begin{references}
\bibitem{14} Earlier Binder cumulant data \cite{4} which gave the 
estimate $T_g = 2.06 \pm 0.02$ were obtained on smaller samples and 
were probably affected by corrections to finite size scaling.
\end{references}
\end{document}